\documentclass[a4paper]{jpconf}
\usepackage{amsmath, amssymb}
\usepackage{mathrsfs}
\usepackage{graphicx}
\usepackage{dcolumn}
\usepackage{bm}
\usepackage{color}
\usepackage{url}

\def\ve#1{{\bm{#1}}}

\def\urm#1{\scriptstyle{\text{\textrm{\textmd{\textup{#1}}}}}}

\let\temp\epsilon
\let\epsilon\varepsilon
\let\varepsilon\temp
\let\temp\relax
\def\defeq{\mathrel{:=}}

\begin{document}
\title{How to Improve Functionals in Density Functional Theory?
  ---Formalism and Benchmark Calculation---}
\author{
  Tomoya Naito$ {}^{1, \, 2} $,
  Daisuke Ohashi$ {}^{1, \, 2} $,
  and
  Haozhao Liang$ {}^{2, \, 1} $}
\address{$ {}^1 $ Department of Physics, Graduate School of Science, The University of Tokyo,
  Tokyo 113-0033, Japan}
\address{$ {}^2 $ RIKEN Nishina Center, Wako 351-0198, Japan}
\ead{naito@cms.phys.s.u-tokyo.ac.jp}
\begin{abstract}
  We proposed in Ref.~[arXiv:1812.09285v2] a way to improve energy density functionals in the density functional theory based on the combination of the inverse Kohn-Sham method and the density functional perturbation theory.
  In this proceeding, we mainly focus on the results for the $ \mathrm{Ar} $ and $ \mathrm{Kr} $ atoms.
\end{abstract}
%
\section{Introduction}
\label{sec:intr}
\par
The density functional theory (DFT) is one of the most successful approaches to
the calculation of the ground-state properties of the quantum many-body problems
including nuclear systems 
\cite{PhysRev.136.B864,PhysRev.140.A1133,Bender2003Rev.Mod.Phys.75_121,RevModPhys.88.045004}.
In principle, the DFT gives the exact ground-state density $ \rho_{\urm{gs}} $ and energy $ E_{\urm{gs}} $:
\begin{equation}
  \label{eq:GSenergy_def}
  E_{\urm{gs}}
  =
  T_0 \left[ \rho_{\urm{gs}} \right]
  +
  \int
  V_{\urm{ext}}
  \left( \ve{r} \right) \,
  \rho_{\urm{gs}} \left( \ve{r} \right) \,
  d \ve{r}
  +
  E_{\urm{H}} \left[ \rho_{\urm{gs}} \right]
  +
  E_{\urm{xc}} \left[ \rho_{\urm{gs}} \right],
\end{equation}
where $ T_0 $ is the Kohn-Sham (KS) kinetic energy,
$ V_{\urm{ext}} $ is the external field,
and $ E_{\urm{H}} \left[ \rho \right] $ and
$ E_{\urm{xc}} \left[ \rho \right] $ are the Hartree and exchange-correlation energy density functionals (EDFs), respectively \cite{PhysRev.136.B864,PhysRev.140.A1133}.
However, in practice, the accuracy of the DFT calculation depends on that of the approximations for $ E_{\urm{xc}} \left[ \rho \right] $,
as it is unknown.
Hence, the derivation or construction of accurate EDFs is one of the primary goals in DFT for both electron and nuclear systems.
In Ref.~\cite{Naito2018_arXiv1812.09285}, we proposed a novel way to improve EDFs based on the combination of the inverse Kohn-Sham (IKS) method \cite{PhysRevA.47.R1591,doi:10.1063/1.465093}
and the density functional perturbation theory (DFPT) \cite{PhysRevLett.58.1861,PhysRevA.52.1096,Gonze1989Phys.Rev.B39_13120,RevModPhys.73.515},
the so-called IKS-DFPT method.
In this method, the first-order DFPT, also called the Hellmann-Feynman theorem \cite{Feynman1939Phys.Rev.56_340}, is used,
and the known functional is improved by using the IKS-DFPT.
As benchmark calculations, we verify this method by reproducing the exchange functional in the local density approximation (LDA) \cite{Proc.Camb.Phil.Soc.26.376}.
In this proceeding, we mainly focus on the results for the $ \mathrm{Ar} $ and $ \mathrm{Kr} $ atoms.
%
\section{Theoretical Framework}
\label{sec:theo}
\par
In the DFT, $ \rho_{\urm{gs}} \left( \ve{r} \right) $ and $ E_{\urm{gs}} $ of an $ N $-particle system are obtained by solving the KS equations,
\begin{equation}
  \label{eq:KSeq}
  \left[
    -\frac{\hbar^{2}}{2m}
    \nabla^{2}
    +
    V_{\urm{KS}} \left( \ve{r} \right)
  \right]
  \psi_i \left( \ve{r} \right)
  =
  \epsilon_i
  \psi_i \left( \ve{r} \right),
  \qquad
  V_{\urm{KS}} \left( \ve{r} \right)
  =
  V_{\urm{ext}} \left( \ve{r} \right)
  +
  \frac{\delta E_{\urm{Hxc}} \left[ \rho_{\urm{gs}} \right]}{\delta \rho \left( \ve{r} \right)},
\end{equation}
where $ m $ is the mass of particles,
$ \psi_i \left( \ve{r} \right) $ and $ \epsilon_i $ are the single-particle orbitals and energies,
respectively,
$ V_{\urm{KS}} \left( \ve{r} \right) $ is the KS effective potential,
and $ \rho_{\urm{gs}} \left( \ve{r}\right) = 
\sum_{i=1}^N
\left|
  \psi_i \left( \ve{r} \right)
\right|^{2} $.
\par
The IKS provides $ V_{\urm{KS}} $ for each system from the ground-state density $ \rho_{\urm{gs}} $,
which can be determined from experiments or high-accuracy calculations,
such as the coupled cluster and the configuration interaction methods for atoms and light molecules
and several \textit{ab initio} methods for light nuclei.
As mentioned in Ref.~\cite{Kohn1999Rev.Mod.Phys.71_1253},
the KS potential $ V_{\urm{KS}} \left( \ve{r} \right) $ is unique concerning the system.
Furthermore, improvement of the EDFs by using the IKS is promising
since the EDF is, in principle, unique for all the electron systems,
In our novel method IKS-DFPT \cite{Naito2018_arXiv1812.09285},
the conventional Hartree-exchange-correlation functional $ \tilde{E}_{\urm{Hxc}} $
will be improved by using the IKS.
\par
Here,
$ \tilde{E}_{\urm{Hxc}} \left[ \rho \right] $ is assumed to be close enough to the exact one
$ E^{\urm{exact}}_{\urm{Hxc}} \left[ \rho \right]$,
since $ \tilde{E}_{\urm{Hxc}} \left[ \rho \right] $ is known to work well.
Hence, the difference between
$ E_{\urm{Hxc}}^{\urm{exact}} $ and $ \tilde{E}_{\urm{Hxc}} $ 
is treated as a perturbation.
If the difference is not small enough to be treated as the perturbation,
the final results would be unreasonable.
\par
The first-order perturbation theory is used for the treatment of the difference between $ E_{\urm{Hxc}}^{\urm{exact}} $ and $ \tilde{E}_{\urm{Hxc}} $ as
\begin{equation}
  \label{eq:idea_Hxc}
  E^{\urm{exact}}_{\urm{Hxc}} \left[ \rho \right]
  =
  \tilde{E}_{\urm{Hxc}} \left[ \rho \right]
  +
  \lambda E^{(1)}_{\urm{Hxc}} \left[ \rho \right]
  +
  O \left( \lambda^{2} \right),
\end{equation}
with a small parameter $ \lambda $.
Then, the exact single-particle orbitals $ \psi^{\urm{exact}}_i \left( \ve{r} \right) $, 
ground-state density $ \rho^{\urm{exact}}_{\urm{gs}} \left( \ve{r} \right) $,
and energy $ E^{\urm{exact}}_{\urm{gs}} $ are also expanded perturbatively:
\numparts
\begin{align}
  \psi^{\urm{exact}}_i \left( \ve{r} \right)
  & =
    \tilde{\psi}_i \left( \ve{r} \right)
    +
    \lambda \psi^{(1)}_i \left( \ve{r} \right)
    +
    O \left( \lambda^{2} \right),
    \label{eq:idea_psi} \\
  \rho^{\urm{exact}}_{\urm{gs}} \left( \ve{r} \right)
  & =
    \tilde{\rho}_{\urm{gs}} \left( \ve{r} \right)
    +
    \lambda \rho_{\urm{gs}}^{(1)} \left( \ve{r} \right)
    +
    O \left( \lambda^{2} \right),
    \label{eq:idea_rho} \\
  E^{\urm{exact}}_{\urm{gs}}
  & =
    \tilde{E}_{\urm{gs}} +\lambda E_{\urm{gs}}^{(1)}
    +
    O \left( \lambda^{2} \right),
    \label{eq:idea_GSenergy}
\end{align}
\endnumparts
where quantities shown with the tilde are given by $ \tilde{E}_{\urm{Hxc}} $.
The first-order perturbation term $ \psi_i^{(1)} \left( \ve{r} \right) $ is assumed to be orthogonal to $ \tilde{\psi}_i \left( \ve{r} \right) $.
The perturbation is assumed not to affect the external field,
i.e.,~$ V^{\urm{exact}}_{\urm{ext}} \left( \ve{r} \right) = \tilde{V}_{\urm{ext}} \left( \ve{r} \right) $.
Moreover, $ \rho^{\urm{exact}}_{\urm{gs}} \left( \ve{r} \right) $ is assumed to be given,
and thus $ \psi^{\urm{exact}}_i \left( \ve{r} \right) $ is calculated from the IKS.
\par
Under these assumptions, we calculate $ E^{\urm{exact}}_{\urm{gs}} $ in two different ways.
One way is based on the first-order DFPT, and the other way is based on the IKS and KS equations.
In the former way, substitution of Eqs.~\eqref{eq:idea_Hxc}, \eqref{eq:idea_psi}, and \eqref{eq:idea_rho} into Eq.~\eqref{eq:GSenergy_def} gives
\fl
\begin{align}
  E_{\urm{gs}}^{\urm{exact}}
  = & \,
      T_0 \left[ \rho_{\urm{gs}}^{\urm{exact}} \right]
      +
      \int
      V_{\urm{ext}} \left( \ve{r} \right) \,
      \tilde{\rho}_{\urm{gs}} \left( \ve{r} \right) \,
      d \ve{r}
      +
      \tilde{E}_{\urm{Hxc}} \left[ \tilde{\rho}_{\urm{gs}} \right]
      +
      \lambda
      E_{\urm{Hxc}}^{(1)} \left[ \tilde{\rho}_{\urm{gs}} \right]
      \notag \\
    & +
      \lambda
      \int
      V_{\urm{ext}} \left( \ve{r} \right) \,
      \rho_{\urm{gs}}^{(1)} \left( \ve{r} \right) \,
      d \ve{r}
      +
      \lambda
      \int
      \frac{\delta \tilde{E}_{\urm{Hxc}} \left[ \tilde{\rho}_{\urm{gs}} \right]}{\delta \rho \left( \ve{r} \right)}
      \rho_{\urm{gs}}^{(1)} \left( \ve{r} \right) \,
      d \ve{r} 
      +
      O \left( \lambda^2 \right).
      \label{eq:gs_1}
\end{align}
In the latter way, Eq.~\eqref{eq:idea_Hxc} and integration of the KS equation \eqref{eq:KSeq} give
\begin{align}
  E^{\urm{exact}}_{\urm{gs}}
  = & \,
      \sum_{i=1}^N
      \epsilon_i^{\urm{exact}}
      +
      E^{\urm{exact}}_{\urm{Hxc}} \left[ \rho^{\urm{exact}}_{\urm{gs}} \right]
      -
      \int
      \frac{\delta E^{\urm{exact}}_{\urm{Hxc}} \left[ \rho^{\urm{exact}}_{\urm{gs}} \right]}
      {\delta \rho \left(\ve{r} \right)}
      \rho^{\urm{exact}}_{\urm{gs}} \left( \ve{r} \right)
      \, d\ve{r}
      \notag \\
  = & \,
      \sum_{i=1}^N
      \epsilon_i^{\urm{exact}}
      +
      \tilde{E}_{\urm{Hxc}} \left[ \rho^{\urm{exact}}_{\urm{gs}} \right]
      +
      \lambda
      E^{(1)}_{\urm{Hxc}} \left[ \rho^{\urm{exact}}_{\urm{gs}} \right]
      \notag \\
    & -
      \int
      \frac{\delta \tilde{E}_{\urm{Hxc}} \left[ \rho^{\urm{exact}}_{\urm{gs}} \right]}
      {\delta \rho \left(\ve{r} \right)}
      \rho^{\urm{exact}}_{\urm{gs}} \left( \ve{r} \right)
      \, d\ve{r}
      -
      \lambda
      \int
      \frac{\delta E^{(1)}_{\urm{Hxc}} \left[ \rho^{\urm{exact}}_{\urm{gs}} \right]}
      {\delta \rho \left(\ve{r} \right)}
      \rho^{\urm{exact}}_{\urm{gs}} \left( \ve{r} \right)
      \, d\ve{r}
      +
      O \left( \lambda^2 \right),
      \label{eq:gsEnergy_2nd}
\end{align}
where $ \epsilon_i^{\urm{exact}} $ are obtained from $ \rho^{\urm{exact}}_{\urm{gs}} $ by using the IKS.
By comparing these two expressions of the ground-state energy
and neglecting the $ O \left( \lambda^2 \right) $ term, 
the equation for $ E^{(1)}_{\urm{Hxc}} \left[ \rho \right] $ is obtained:
\begin{align}
  & \lambda E^{(1)}_{\urm{Hxc}} \left[ \tilde{\rho}_{\urm{gs}} \right]
    -
    \lambda E^{(1)}_{\urm{Hxc}} \left[ \rho^{\urm{exact}}_{\urm{gs}} \right]
    +
    \lambda
    \int
    \frac{\delta E^{(1)}_{\urm{Hxc}} \left[ \rho^{\urm{exact}}_{\urm{gs}} \right]}
    {\delta \rho \left( \ve{r} \right)}
    \rho^{\urm{exact}}_{\urm{gs}} \left( \ve{r} \right)
    \, d \ve{r}
    \notag \\
  = & \, 
      \sum_{i=1}^N \epsilon_i^{\urm{exact}}
      +
      \tilde{E}_{\urm{Hxc}} \left[ \rho^{\urm{exact}}_{\urm{gs}} \right]
      -
      \int
      \frac{\delta \tilde{E}_{\urm{Hxc}} \left[ \rho^{\urm{exact}}_{\urm{gs}} \right]}
      {\delta \rho \left( \ve{r} \right)}
      \rho^{\urm{exact}}_{\urm{gs}} \left( \ve{r} \right)
      \, d \ve{r}
      -
      \tilde{E}_{\urm{gs}}
      \defeq 
      C \left[ \rho^{\urm{exact}}_{\urm{gs}} \right].
      \label{eq:basic_eq}
\end{align}
The right-hand side of this equation can be calculated from the known quantities
and
its value depends only on the exact ground-state density $ \rho_{\urm{gs}}^{\urm{exact}} $ 
and the known functional $ \tilde{E}_{\urm{Hxc}} $.
Thus, hereafter the right-hand side of the equation is shown as $ C \left[ \rho \right] $.
\par
Finally, solving Eq.~\eqref{eq:basic_eq},
the Hartree-exchange-correlation functional in the IKS-DFPT in the first-order,
i.e.,~the IKS-DFPT1, is derived as 
\begin{equation}
  \label{eq:calc}
  E_{\urm{Hxc}} \left[ \rho \right]
  =
  \tilde{E}_{\urm{Hxc}} \left[ \rho \right]
  +
  \lambda E^{(1)}_{\urm{Hxc}} \left[ \rho \right].
\end{equation}
\par
Because Eq.~\eqref{eq:calc} is a functional equation, it is difficult to be solved directly.
In this work, we assume
\begin{equation}
  E^{(1)}_{\urm{Hxc}} \left[ \rho \right]
  =
  A
  \int
  \left[
    \rho\left(\ve{r}\right)
  \right]^{\alpha}
  \, d \ve{r},
  \label{eq:PC_new}
\end{equation}
with the values of $ A $ and $ \alpha $ to be determined, and then we get
\begin{equation}
  \label{eq:final_eq}
  \lambda
  A \int
  \left\{
    \left[
      \tilde{\rho}_{\urm{gs}} \left( \ve{r} \right)
    \right]^{\alpha}
    +
    \left( \alpha - 1 \right)
    \left[ \rho^{\urm{exact}}_{\urm{gs}} \left( \ve{r} \right) \right]^{\alpha}
  \right\}
  \, d \ve{r}
  =
  C \left[ \rho^{\urm{exact}}_{\urm{gs}} \right].
\end{equation}
To determine $ A $ and $ \alpha $, two systems, Systems 1 and 2, are required.
Here, $ \rho_1 $ and $ \rho_2 $ are the exact ground-state densities,
and $ \tilde{\rho}_1 $ and $ \tilde{\rho}_2 $ are the ground-state densities 
of Systems 1 and 2 calculated from $ \tilde{E}_{\urm{Hxc}} \left[ \rho \right] $, respectively.
Substituting $ \rho_i $ and $ \tilde{\rho}_i $ ($ i = 1, \, 2 $) for Eq.~\eqref{eq:basic_eq},
it leads to the two equations for $ \lambda A $ and $ \alpha $.
In such a way, $ \lambda A $ and $ \alpha $ can be determined.
Note that in principle the Hartree-exchange-correlation EDF is system independent,
and thus any system can be used as Systems 1 and 2.
\section{Benchmark Calculations and Discussions}
\label{sec:calc}
\par
As benchmark calculations,
to avoid ambiguity coming from the experimental data, 
we use $ \rho^{\urm{target}}_{\urm{gs}} \left( \ve{r} \right) $ 
calculated from the theoretical $ E^{\urm{target}}_{\urm{Hxc}} \left[ \rho \right] $ 
as $ \rho^{\urm{exact}}_{\urm{gs}} \left( \ve{r} \right) $,
and we test whether $ E^{\urm{target}}_{\urm{Hxc}} \left[ \rho \right] $ can be reproduced from $ \tilde{E}_{\urm{Hxc}} \left[ \rho \right] $ in this scheme.
The Hartree and 
the Hartree plus LDA exchange functional (Hartree-Fock-Slater approximation) \cite{Proc.Camb.Phil.Soc.26.376}
are used for 
$ \tilde{E}_{\urm{Hxc}} \left[ \rho \right] $ and 
$ E^{\urm{target}}_{\urm{Hxc}} \left[ \rho \right] $, respectively,
as a benchmark:
\begin{equation}
  \label{eq:1stCase}
  \tilde{E}_{\urm{Hxc}} \left[ \rho \right]
  = 
  \frac{1}{2}
  \iint
  \frac{\rho \left( \ve{r} \right) \, \rho \left( \ve{r}' \right)}
  {\left| \ve{r} - \ve{r}' \right|}
  \, d \ve{r}
  \, d \ve{r}', 
  \quad 
  E^{\urm{target}}_{\urm{Hxc}} \left[ \rho \right]
  =
  \tilde{E}_{\urm{Hxc}} \left[ \rho \right]
  -
  \frac{3}{4}
  \left(
    \frac{3}{\pi}
  \right)^{1/3}
  \int
  \left[
    \rho \left( \ve{r} \right)
  \right]^{4/3}
  \, d \ve{r}
\end{equation}
in the Hartree atomic unit.
All the pairs of the isolated noble-gas atoms ($ \mathrm{He} $, $ \mathrm{Ne} $, $ \mathrm{Ar} $, $ \mathrm{Kr} $, $ \mathrm{Xe} $, and $ \mathrm{Rn} $) are used as Systems 1 and 2.
The external field
$ V^{\urm{target}}_{\urm{ext}} \left( \ve{r} \right) = \tilde{V}_{\urm{ext}} \left( \ve{r} \right) $
is the Coulomb interaction between the nucleus and electron.
\par
In Table~\ref{tab:it_hx_ArKr},
the coefficients $ \alpha $ and $ \lambda A $ and the ground-state energies $ E_{\urm{gs}} $ calculated in the IKS-DFPT are shown for the pair of atoms $ \mathrm{Ar} $-$ \mathrm{Kr} $.
It is found that $ \alpha $ and $ \lambda A $ are obtained within $ 0.3 \, \% $ and $ 3.7 \, \% $ errors from their target values, respectively.
In Table~\ref{tab:systematic},
the coefficients calculated in all the pairs and their errors with respect to the target valued are shown.
Among all the pairs, $ \alpha $ is obtained within more or less $ 1.0 \, \% $ errors.
In contrast, $ \lambda $ is obtained with around $ 5 \, \% $ errors.
Both coefficients calculated from the heavier atoms are more accurate.
This comes from the fact that the density of heavier atom ranges wider than that of lighter atom.
\par
\begin{figure}[t]
  \centering
  \includegraphics[width=8cm]{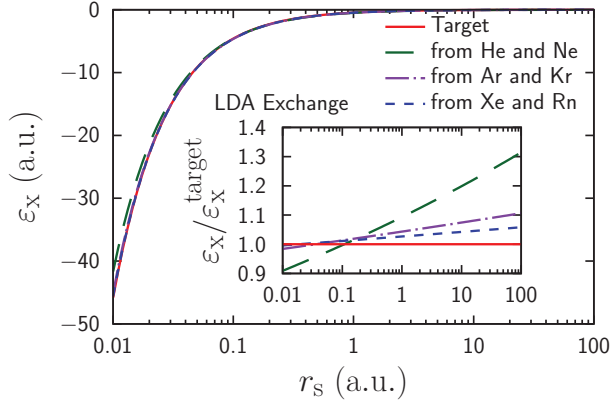}
  \caption{
    Energy density $ \epsilon_{\urm{x}} $ for the LDA exchange functional as a function of $ r_{\urm{s}} $.
    Ratios of $ \epsilon_{\urm{x}} / \epsilon^{\urm{target}}_{\urm{x}} $ are shown in the insert.}
  \label{fig:hx_all}
\end{figure}

The calculated exchange energy density
$ \epsilon_{\urm{x}} \left( r_{\urm{s}} \right) $
and the ratios to the target one
$ \epsilon_{\urm{x}} \left( r_{\urm{s}} \right) / \epsilon^{\urm{target}}_{\urm{x}} \left( r_{\urm{s}} \right) $
are shown as a function of $ r_{\urm{s}} $ in Fig.~\ref{fig:hx_all}
for the pairs of $ \mathrm{He} $-$ \mathrm{Ne} $, $ \mathrm{Ar} $-$ \mathrm{Kr} $, and $ \mathrm{Xe} $-$ \mathrm{Rn} $ with the dashed, dot-dashed, and dotted lines, respectively,
while the target one is shown with the solid line.
Here, the energy density $ \epsilon_{\urm{x}} \left( \rho \right) $
and the Wigner-Seitz radius $ r_{\urm{s}} $ are defined as
$ E_{\urm{x}} \left[ \rho \right] =
\int
\epsilon_{\urm{x}} \left( \rho \right) \,
\rho \left( \ve{r} \right)
\, d \ve{r} $ and 
$ r_{\urm{s}} =
\left[
  3 / \left(4 \pi \rho \right)
\right]^{1/3} $, respectively.
The pair of $ \mathrm{Xe} $-$ \mathrm{Rn} $ reproduces the target functional within a few percents in the range of $ 0.01 \, \mathrm{a.u.} \le r_{\urm{s}} \le 100 \, \mathrm{a.u.} $,
which is generally better than the pair of $ \mathrm{He} $-$ \mathrm{Ne} $.
As comparing 
$ \mathrm{He} $-$ \mathrm{Ne} $, $ \mathrm{Ar} $-$ \mathrm{Kr} $, and $ \mathrm{Xe} $-$ \mathrm{Rn} $ cases,
better reproduction in the high-density region leads to better reproduction of the coefficients,
since the polynomial form of the functional in Eq.~\eqref{eq:PC_new} is more sensitive to the high-density region.
\par
\begin{figure}[t]
  \centering
  \includegraphics[width=8cm]{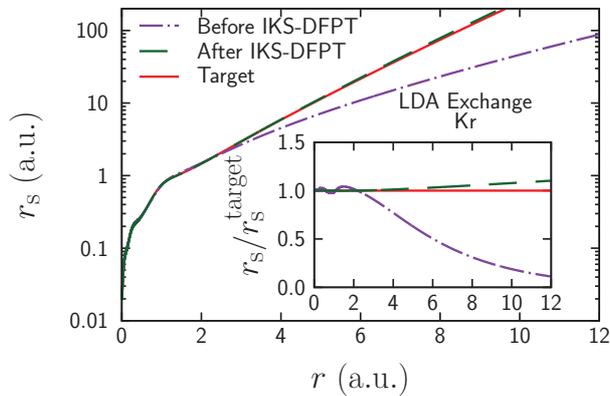}
  \caption{
    Wigner-Seitz radii $ r_{\urm{s}} $ as a function of $ r $ for $ \mathrm{Kr} $.
    Ratios of $ r_{\urm{s}} / r^{\urm{target}}_{\urm{s}} $ are shown in the insert.}
  \label{fig:it_density_Kr_hx}
\end{figure}

The Wigner-Seitz radii $ r_{\urm{s}} $ calculated in the functional before and after the IKS-DFPT 
and the target one $ r_{\urm{s}}^{\urm{target}} $ for $ \mathrm{Kr} $ 
are shown as a functions of $ r $ in Fig.~\ref{fig:it_density_Kr_hx}
with the dot-dashed, dashed, and solid lines, respectively.
The ratios of calculated Wigner-Seitz radius to the target one, 
$ r_{\urm{s}} / r^{\urm{target}}_{\urm{s}} $,
are also shown in the insert of Fig.~\ref{fig:it_density_Kr_hx}.
It is found that the ground-state density is also much improved after the IKS-DFPT is performed.
\begin{table}[t]
  \caption{
    Coefficients $ \alpha $ and $ \lambda A $ and the ground-state energies $ E_{\urm{gs}} $ calculated in the IKS-DFPT 
    for the pair of atoms $ \mathrm{Ar} $ and $ \mathrm{Kr} $.
    The Hartree functional is used for $ \tilde{E}_{\urm{Hxc}}\left[\rho\right] $
    and the Hartree plus LDA exchange functional is used for $ E^{\urm{target}}_{\urm{Hxc}} \left[ \rho \right] $
    given in Eq.~\eqref{eq:1stCase}.
    All units are in the Hartree atomic unit.}
  \label{tab:it_hx_ArKr}
  \centering
  \begin{tabular}{lrrrr}
    \hline \hline
    & \multicolumn{1}{c}{$ \alpha $} & \multicolumn{1}{c}{$ \lambda A $} & \multicolumn{1}{c}{$ E_{\urm{gs}}$ of $ \mathrm{Ar} $} & \multicolumn{1}{c}{$ E_{\urm{gs}}$ of $ \mathrm{Kr} $} \\ \hline
    Original (Hartree) & & & $ -497.3858 $ & $ -2659.6912 $ \\
    IKS-DFPT & $ 1.3290958 $ & $ -0.7658732 $ & $ -525.1119 $ & $ -2748.1434 $ \\ \hline
    Target (Hartree-Fock-Slater) & $ 1.3333333 $ & $ -0.7385588 $ & $ -524.5143 $ & $ -2746.7828 $ \\ 
    \hline \hline
  \end{tabular}
\end{table}
\begin{table}[t]
  \centering
  \caption{
    Coefficients $ \alpha $ and $ \lambda A $ for all the pairs of noble-gas atoms.
    The errors with respect to the target values are also shown.
    All units are in the Hartree atomic unit.}
  \label{tab:systematic}
  \begin{tabular}{lD{.}{.}{7}D{.}{.}{6}D{.}{.}{7}D{.}{.}{6}}
    \hline \hline
    Pairs & \multicolumn{1}{c}{Exchange $ \alpha $} & \multicolumn{1}{c}{Error for $ \alpha $ ($ \mathrm{\%} $)} & \multicolumn{1}{c}{Exchange $ \lambda A $} & \multicolumn{1}{c}{Error for $ \lambda A $ ($ \mathrm{\%} $)} \\ \hline
    Target                          & 1.3333333 & \multicolumn{1}{c}{---} & -0.7385588 & \multicolumn{1}{c}{---} \\ \hline
    $ \mathrm{He} $-$ \mathrm{Ne} $ & 1.3199872 & 1.000960 & -0.7920448 & 7.241947 \\
    $ \mathrm{He} $-$ \mathrm{Ar} $ & 1.3209765 & 0.926762 & -0.7926638 & 7.325759 \\
    $ \mathrm{Ne} $-$ \mathrm{Ar} $ & 1.3235352 & 0.734860 & -0.7841588 & 6.174192 \\
    $ \mathrm{He} $-$ \mathrm{Kr} $ & 1.3227758 & 0.791815 & -0.7937863 & 7.477744 \\
    $ \mathrm{Ne} $-$ \mathrm{Kr} $ & 1.3263436 & 0.524230 & -0.7779323 & 5.331131 \\
    $ \mathrm{Ar} $-$ \mathrm{Kr} $ & 1.3290958 & 0.317815 & -0.7658732 & 3.698343 \\
    $ \mathrm{He} $-$ \mathrm{Xe} $ & 1.3235844 & 0.731170 & -0.7942892 & 7.545836 \\
    $ \mathrm{Ne} $-$ \mathrm{Xe} $ & 1.3270817 & 0.468872 & -0.7762984 & 5.109903 \\
    $ \mathrm{Ar} $-$ \mathrm{Xe} $ & 1.3292187 & 0.308597 & -0.7654719 & 3.644007 \\
    $ \mathrm{Kr} $-$ \mathrm{Xe} $ & 1.3294148 & 0.293890 & -0.7644846 & 3.510328 \\
    $ \mathrm{He} $-$ \mathrm{Rn} $ & 1.3244450 & 0.666625 & -0.7948236 & 7.618193 \\
    $ \mathrm{Ne} $-$ \mathrm{Rn} $ & 1.3279028 & 0.407290 & -0.7744818 & 4.863937 \\
    $ \mathrm{Ar} $-$ \mathrm{Rn} $ & 1.3297748 & 0.266890 & -0.7636589 & 3.398529 \\
    $ \mathrm{Kr} $-$ \mathrm{Rn} $ & 1.3303022 & 0.227335 & -0.7606336 & 2.988907 \\
    $ \mathrm{Xe} $-$ \mathrm{Rn} $ & 1.3311445 & 0.164162 & -0.7558229 & 2.337544 \\ \hline
    Average                         & 1.3263722 & 0.522085 & -0.7770949 & 5.217753 \\ 
    \hline \hline
  \end{tabular}
\end{table}
%
\section{Conclusion and Perspectives}
\label{sec:conc}
\par
In summary, the way to improve conventional EDFs based on the combination of the IKS and the DFPT was proposed in Ref.~\cite{Naito2018_arXiv1812.09285}.
As benchmark calculations, we test whether the LDA exchange functional can be reproduced in this novel scheme IKS-DFPT1.
By improving the exchange functional, the accuracy of the ground-state energies is improved by two to three orders of magnitude,
and the accuracy of the ground-state densities is also improved one to two orders of magnitude.
Therefore, the IKS-DFPT is promising to improve the conventional functionals.
Application of this IKS-DFPT to the nuclear DFT is promising.
\ack
\par
T.N.~and D.O.~acknowledge the financial support from Computational Science Alliance, The University of Tokyo.
T.N.~and H.L.~would like to thank the RIKEN iTHEMS program
and the JSPS-NSFC Bilateral Program for Joint Research Project on Nuclear mass and life for unravelling mysteries of the $ r $-process.
T.N.~acknowledges the JSPS Grant-in-Aid for JSPS Fellows under Grant No.~19J20543.
H.L.~acknowledges the JSPS Grant-in-Aid for Early-Career Scientists under Grant No.~18K13549.
\section*{References}
\bibliographystyle{iopart-num}
\bibliography{citation}

\providecommand{\newblock}{}
\begin{thebibliography}{10}
\expandafter\ifx\csname url\endcsname\relax
  \def\url#1{{\tt #1}}\fi
\expandafter\ifx\csname urlprefix\endcsname\relax\def\urlprefix{URL }\fi
\providecommand{\eprint}[2][]{\url{#2}}

\bibitem{PhysRev.136.B864}
Hohenberg P and Kohn W 1964 {\em Phys.~Rev.\/} {\bf 136} B864

\bibitem{PhysRev.140.A1133}
Kohn W and Sham L~J 1965 {\em Phys.~Rev.\/} {\bf 140} A1133

\bibitem{Bender2003Rev.Mod.Phys.75_121}
Bender M, Heenen P~H and Reinhard P~G 2003 {\em Rev.~Mod.~Phys.\/} {\bf 75} 121

\bibitem{RevModPhys.88.045004}
Nakatsukasa T, Matsuyanagi K, Matsuo M and Yabana K 2016 {\em
  Rev.~Mod.~Phys.\/} {\bf 88} 045004

\bibitem{Naito2018_arXiv1812.09285}
Naito T, Ohashi D and Liang H 2018 {Improvement of Functionals in Density
  Functional Theory by the Inverse Kohn-Sham Method and Density Functional
  Perturbation Theory} (\textit{Preprint} \eprint{arXiv: 1812.09285v2})

\bibitem{PhysRevA.47.R1591}
Wang Y and Parr R~G 1993 {\em Phys.~Rev.~A\/} {\bf 47} R1591

\bibitem{doi:10.1063/1.465093}
Zhao Q and Parr R~G 1993 {\em J.~Chem.~Phys.\/} {\bf 98} 543

\bibitem{PhysRevLett.58.1861}
Baroni S, Giannozzi P and Testa A 1987 {\em Phys.~Rev.~Lett.\/} {\bf 58} 1861

\bibitem{PhysRevA.52.1096}
Gonze X 1995 {\em Phys.~Rev.~A\/} {\bf 52} 1096

\bibitem{Gonze1989Phys.Rev.B39_13120}
Gonze X and Vigneron J~P 1989 {\em Phys.~Rev.~B\/} {\bf 39} 13120

\bibitem{RevModPhys.73.515}
Baroni S, de~Gironcoli S, Dal~Corso A and Giannozzi P 2001 {\em
  Rev.~Mod.~Phys.\/} {\bf 73} 515

\bibitem{Feynman1939Phys.Rev.56_340}
Feynman R~P 1939 {\em Phys.~Rev.\/} {\bf 56} 340

\bibitem{Proc.Camb.Phil.Soc.26.376}
Dirac P~A~M 1930 {\em Proc.~Camb.~Phil.~Soc.\/} {\bf 26} 376

\bibitem{Kohn1999Rev.Mod.Phys.71_1253}
Kohn W 1999 {\em Rev.~Mod.~Phys.\/} {\bf 71} 1253

\end{thebibliography}
\end{document}